\documentclass[twoside,twocolumn,10pt]{article}
\usepackage{wscg}           % includes a number of other packages (e.g., myalgorithm)
\RequirePackage{ifpdf}
\ifpdf
 \RequirePackage[pdftex]{graphicx}
 \RequirePackage[pdftex]{color}
\else
 \RequirePackage[dvips,draft]{graphicx}
 \RequirePackage[dvips]{color}
\fi
\def\le{\left(}
\def\ri{\right)}
\def\dis{\displaystyle}

\usepackage{nopageno}       % no page numbers at all; uncomment for final version

\title{DGLAP-BFKL duality from QCD to quantum computers}

\author{
\parbox{0.25\textwidth}{\centering
Igor Kondrashuk\\[1mm]
GMA {\rm \&} GFAE  {\rm \&}  CCE,  Departamento  de Ciencias  B\'asicas,  
  Universidad del B\'io-B\'io, Campus Fernando May, Av. Andres Bello 720, 
3780227,         Chill\'an, Chile \\ [1mm]
igor.kondrashuk@gmail.com
}
}

\usepackage{url}
\makeatletter
\def\Uslash{\mathbin{\mathchar`\/}\@ifnextchar{/}{\kern-.15em}{}}
\g@addto@macro\UrlSpecials{\do \/ {\Uslash}}
\def\Ucolon{\mathbin{\mathchar`:}\@ifnextchar{/}{\kern-.1em}{}}
\g@addto@macro\UrlSpecials{\do : {\Ucolon}}
\makeatother

\begin{document}

\twocolumn[{\csname @twocolumnfalse\endcsname

\maketitle  % full width title

\begin{abstract}
\noindent
DGLAP integro-differential equation can be solved by applying certain map in the complex plane of Mellin moments. It may be re-written as Schr\"odinger equation for $n$ particles. By applying another map in the complex plane of Mellin moment we may re-write the  DGLAP equation as a dual DGLAP equation.
Regge limit of the dual DGLAP equation coincides with BFKL integro-differential equation which in turn is the Regge limit of optic theorem 
in quantum field theory and may be re-written as another Schr\"odinger equation. This means that the BFKL equation and the corresponding Schr\"odinger equation may be solved by the proposed method of complex mapping in the complex plane of Mellin moments. This approach may be useful in solving tasks related to quantum communication processes.   
\end{abstract}

\subsection*{Keywords}
DGLAP equation, Schr\"odinger equation, complex mapping, BFKL equation, optic theorem, quantum communication process. 

\vspace*{1.0\baselineskip}
}]

%%%%%%%%%%%%%%%%%%%%%%%%%%%%%%%%%%%%%%%%%%%%%%%%%%%%%%%%%%%%%%%%%%%%%%%%%%%%%

\section{Introduction}

\copyrightspace
Two integro-differential equations (IDEs) namely Dokshitser-Gribov-Lipatov-Altarelli-Parisi (DGLAP)  equation \cite{Gribov:1972ri,Gribov:1972rt,Lipatov:1974qm,Dokshitzer:1977sg,Altarelli:1977zs} and Balitsky-Fadin-Kuraev-Lipatov equation (BFKL) \cite{Lipatov:1976zz,Fadin:1975cb,Kuraev:1976ge,Kuraev:1977fs,Balitsky:1978ic}
play important role in quantum chromodynamics (QCD) which is gauge theory responsible for strong interaction between quarks and gluons inside of protons.  In \cite{Altarelli:2000dw,Ball:2005mj}
    these two IDEs were treated as dual equations in a common kinematic region.   
We have treated this duality via complex diffeomorphism in the plane of complex variable that represents Mellin moments in \cite{Kondrashuk:2019cwi},  Actually, we have treated the BFKL equation as a Regge limit of an IDE dual to DGLAP IDE.   

We proposed in   \cite{Alvarez:2019eaa} a method to solve DGLAP IDE via certain complex maps in the complex plane of the  Mellin moment variable.  For us in this note application of this method to  quantum mechanics is important. The DGLAP IDE may be transformed to Schr\"odinger equation, which may be solved by our methods via complex diffeomorphisms.  In this note we talk about the possible application of our method to quantum mechanics    and to quantum computers. We first pay attention to the solution to this  Schr\"odinger equation which is equivalent to the DGLAP IDE.    
This would allow to solve Schr\"odinger equation for different quantum systems via our method of complex mapping in the plane of Mellin moments
\cite{Alvarez:2019eaa}. 

 Because the BFKL IDE is a Regge limit of the dual DGLAP IDE constructed from the DGLAP IDE by a certain complex map, the same method  of complex maps  may be applied to the BFKL IDE and dual DGLAP IDE written as a Schr\"odinger equation. Both the BFKL IDE and the dual DGLAP IDE 
are related to the optic theorem and may be useful in tasks of quantum communication processes.

\section{DGLAP as Quantum Mecanics}

Gribov and Lipatov used Bethe-Salpeter equation in quantum electrodynamics (QED) to study structure funtions of proton in the framework of a parton model \cite{Gribov:1972ri,Gribov:1972rt,Lipatov:1974qm}.  The integro-differentail equations we are talking about in these notes  appeared in the processes of   deeply ineleastic lepton scattering  in which proton structure functions are measured \cite{Collins:1989gx}.  The pointlike partons were ``seen'' inside the hadron in the experiments related to this scattering. 
Discovering of asymptotic freedom for the running coupling in non-Abelian gauge theories in  combination with operator product expansion 
(OPE) allowed to look at the  proton structure functions and at the parton model under a different angle \cite{Gross:1974cs}. OPE may be applied for large momentum transfers in the scattering processes \cite{Gross:1974cs}.  The running coupling may be solved in a perturbative way for large momentum transfers, too
\cite{Cvetic:2011vy}.

We make a brief review of how the DGLAP equation was discovered.  The details may be found in  \cite{Collins:1989gx}. 
The standard hadronic tensor $W^{\mu\nu} (q^\mu, p^\mu)$,
\begin{eqnarray} \label{W}
  &&W^{\mu\nu}  = \frac{1 }{4\pi} \int d^4y e^{ i q \cdot y}
      \sum_X  \langle A|j ^\mu (y) | X \rangle \langle X|j^\nu(0) | A \rangle
\nonumber\\
  &=& F_1(x, Q^2)  \left( -g^{ \mu\nu} + \frac{q^\mu q^\nu }{q^2} \right)
\nonumber\\
  & +& F_2(x, Q^2) \frac{ \left( p^\mu - q^\mu p \cdot q / q^2 \right)
                         \left( p^\nu - q^\nu p \cdot q / q^2 \right)}{p \cdot q}
\end{eqnarray}
may be extracted from measurements in deeply inelastic lepton scattering process
$e + A \to e + X$, see \cite{Collins:1989gx}, in which  $q^\mu$ is the momentum of
a virtual photon of the exchange, $Q^2 = - q_\mu q^\mu$,
$x = Q^2 / 2 q \cdot p$, $p^\mu$ is the momentum of
the incoming hadron $A$, and $j^\mu(y)$ is the electromagnetic current.  In this note $q^\mu, j^\mu(y), p^\mu$ are vectors in the four-dimensional Minkowski space.

The proton structure functions admit this integral representation
\begin{eqnarray}  
 F_1\left(x, \frac{Q^2}{\Lambda^2}\right)   = \sum_{a}\int\limits_x^1 \frac{d \xi}{\xi}  \ f_{a/A}\left(\xi, \frac{\mu^2}{\Lambda^2} \right)   \nonumber \\
H_{1a} \left( \frac{x}{\xi}, \frac{Q^2}{\mu^2}, \alpha\left(\frac{Q^2}{\mu^2}\right)\right),  \label{F1}  \\ 
  \frac{1 }{x}F_2\left(x, \frac{Q^2}{\Lambda^2}\right)     = \sum_{a}
      \int\limits_x^1 \frac{d \xi}{\xi}  \ f_{a/A}\left(\xi, \frac{\mu^2}{\Lambda^2} \right)  \frac{\xi }{x} \nonumber \\
H_{2a} \left( \frac{x}{\xi}, \frac{Q^2}{\mu^2},  \alpha\left(\frac{Q^2}{\mu^2}\right)\right).  \label{F2}  
\end{eqnarray}
Here $ f_{a/A}\left(\xi, \frac{\mu^2}{\Lambda^2} \right)$ is a parton distribution function (PDF), $\mu^2$ is a renormalization scale, $\Lambda^2$ is 
so-called QCD scale. The PDF $ f_{a/A}\left(\xi, \frac{\mu^2}{\Lambda^2} \right)$
means the probability to find a parton of
type $a $ in a hadron of type $A$ carrying a fraction $\xi$ to $\xi + d \xi$
of the hadron's momentum.  In the formula, one sums over all the possible
types of parton  $a$  \cite{Collins:1989gx}.  The functions $H_{1a},  H_{2a}$  may be calculated 
perturbatively.

The factorization  in the integrands of Eqs. (\ref{F1}) and    (\ref{F2}) has its origin in the factorization of the Mellin moments 
of the proton structure functions, 
\begin{eqnarray}   \label{mome}
   \tilde F_1\left(N, \frac{Q^2}{\Lambda^2}\right)
   &=& \int\limits_0^1 \frac{d x}{x}\ x^N F_1\left(x,\frac{Q^2}{\Lambda^2}\right) , \nonumber\\
   \tilde F_2\left(N, \frac{Q^2}{\Lambda^2}\right)
   &=& \int\limits_0^1 \frac{d x}{x} \ x^{N-1} F_2 \left(x, \frac{Q^2}{\Lambda^2}\right),   \nonumber\\
\tilde f_{a/A}\left(N, \frac{\mu^2}{\Lambda^2} \right)
   &=& \int\limits_0^1 \frac{d x}{x} \ x^{N} f_{a/A}\left(x, \frac{\mu^2}{\Lambda^2} \right),
\end{eqnarray}
which become sums of    products of the Mellin moments,
\begin{eqnarray}
\tilde F_i \left(N, \frac{Q^2}{\Lambda^2}\right) \nonumber\\
= \sum_{a}
 \tilde f_{a/A}\left(N, \frac{\mu^2}{\Lambda^2} \right)
\tilde H_{i a} \!\!\left( N, \frac{Q^2}{\mu^2}, \alpha\left(\frac{Q^2}{\mu^2}\right)\right).   \label{start}
\end{eqnarray}
This product should satisfy the evolution equations  with respect to momentum transfer  $Q^2,$  which may be transformed to the 
renormalization group equation with respect to $Q^2$   for the  perturbative part $\tilde H_{i a} \!\!\left( N, \frac{Q^2}{\mu^2}, \alpha\left(\frac{Q^2}{\mu^2}\right)\right).$   The l.h.s. of 
(\ref{start})  does not depend on the renormalization scale $\mu.$ This means that the renormalization group equation for the perturbative part 
$\tilde H_{i a} \!\!\left( N, \frac{Q^2}{\mu^2}, \alpha\left(\frac{Q^2}{\mu^2}\right)\right)$  of (\ref{start}) may be transformed to the renormalization group equation for the Mellin moments $\tilde f_{a/A}\left(N, \frac{\mu^2}{\Lambda^2} \right)$   of the  parton distribution functions.   Each factor  in (\ref{start}) satisfies  its own evolution  equation with respect to  the renormalization group. However, these evolution equations are related because the lhs does not depend on $\mu.$ In particular, the anomalous dimensions of the Mellin moments of the parton distribution funactions coincide with anomalous dimensions of the operators of highest twist from the OPE  \cite{Altarelli:1977zs}.

Because of  the reasons explained in \cite{Alvarez:2016juq} the dominant  partons  in the region of small $x$ are gluons.  The sum over partons  in (\ref{start}) may be omitted in this approximation  and we leave with a gluon distribution function only, 
 \begin{equation}
\tilde F_i \left(N, \frac{Q^2}{\Lambda^2}\right) \sim   \tilde f \left(N, \frac{\mu^2}{\Lambda^2} \right)
\tilde H_{i} \left( N, \frac{Q^2}{\mu^2}, \alpha\left(\frac{Q^2}{\mu^2}\right)\right).    \label{domi}
\end{equation}

It was shown by  Altarelli and Parisi in \cite{Altarelli:1977zs}  that the evolution equations for the PDF is 
 \begin{eqnarray}  \label{rge}
u \frac{d}{d u} \tilde f(N,u)  = \frac{\alpha(u)}{2\pi}~\gamma(N,\alpha(u))~\tilde f(N,u).
\end{eqnarray}
where we have introduced the notation $u \equiv \mu^2/\Lambda^2,$ and $\gamma(N,\alpha(u))$ is the anomalous dimension of the gluon  distribution function.  The RG equations for the perturbative parts 
$\tilde H_{i} \left( N, \frac{Q^2}{\mu^2}, \alpha\left(\frac{Q^2}{\mu^2}\right)\right)$ 
are more complicate \cite{Gross:1974cs}.

The renormalization group equation for the  Mellin moments of the parton distribution functions can be represented as an IDE for the same parton distribution functions \cite{Altarelli:1977zs}.  After transforming  Mellin moments back to the PDF we obtain  
\begin{eqnarray}  
u\frac{d}{du} f(x,u) =  \frac{\alpha(u)}{2\pi}\int\limits_x^1\frac{dy}{y}~f(y,u)P\left(\frac{x}{y}, \alpha(u) \right),  \label{DGLAP-1} \\
\int\limits_0^1 dx ~x^{N-1} P\left(x, \alpha(u)\right) = \gamma(N,\alpha(u)), \label{split}
\end{eqnarray}
These equations for the PDF are similar to  Gribov and Lipatov IDE equations for the PDF  found in QED based on Bethe-Salpeter  equation with similar  splitting functions   $P\left(\frac{x}{y}, \alpha(u)\right) $ in the integral kernels of these equations  
\cite{Gribov:1972ri,Gribov:1972rt,Lipatov:1974qm}.   The anomalous dimension  $\gamma(N,\alpha(u))$  is the Mellin moment  of the splitting function    $P\left(\frac{x}{y}, \alpha(u)\right). $ The results of Gribov and Lipatov \cite{Gribov:1972ri,Gribov:1972rt,Lipatov:1974qm} were generalized  from QED to QCD by Dokshitzer \cite{Dokshitzer:1977sg}.

The renormalization group equation for the running coupling is solved only  up to some order of the perturbation theory \cite{Cvetic:2011vy}.  The solution of \cite{Cvetic:2011vy} for the running coupling  possesses the asymptotic freedom at large momentum transfers $Q^2.$  As to the region of low momentum transfers $Q^2,$   the running coupling  is not necessary  large or infinite \cite{Ayala:2017tco}.  However, there is one model  \cite{Novikov:1983uc,Jones} of ${\cal N}=1$ supersymmetric pure Yang-Mills  theory with  $SU(N)$ as a gauge group,   in which the running  coupling  may be solved analytically in all the orders of the perturbation theory    \cite{Cvetic:2011vy} . The solution given in   \cite{Cvetic:2011vy} is 
\begin{eqnarray*}
\alpha\left(  \frac{Q^2}{\mu^2} \right)   = -\frac{2\pi}{N} \frac{1}{W_{\mp 1}\left(-(\mu^2/Q^2)^{3/2}\right)}
\end{eqnarray*}
in which $W_{\mp 1}$ are different branches of Lambert function on the complex plane.   In this formula $N$ is from the the gauge group $SU(N)$ of the Yang-Mills theory. In all the other formulas of this paper $N$ is the variable of the Mellin moment.  This solution shows asymptotic freedom at large momentum transfers  $Q^2$    \cite{Cvetic:2011vy}.

The structure of  (\ref{start}), factorized in the product of the perturbative part and the PDF, comes from the solution to the renormalization group equations for products of operators  \cite{Gross:1974cs}.      Initially, at some scale,  the Mellin moment  of the proton structure function 
is taken as  a product of a factor depending on the QCD scale  $\Lambda$  and on the  Mellin variable $N,$  and  the Mellin moment of a coefficient  function \cite{Gross:1974cs}.  The latter may be calculated in a perturbative way \cite{Collins:1989gx}.  The moments of PDFs are parametrized at certain scale in certain way, for example,  by using the Euler beta function \cite{Alekhin:2003ev}.   At another scale they are different and evolve according 
 to  the renormalization group equation.  Recall  that  OPE is valid for the  large momentum transfers, but  at this range of energies there is asymptotic freedom. Asymptotic freedom allows to treat patrons as non interacting particles distributed inside of protons.  Parton model is consistent with  OPE and asymptotic freedom because it is valid for large momentum transfers $Q^2.$

Lipatov in the framework of the parton model introduced  in  \cite{Lipatov:1974qm}  the wave functions  $\Psi\left( \beta_1, k_{\perp 1}; \beta_2, k_{\perp 2};\dots;\beta_n, k_{\perp n}\right)$  of the hadron in its infinite momentum frame $|\bf p| \rightarrow \infty,$  in which $k_{i}$ are 
the parton momenta,  $k_{\perp i}$  are their components transverse to to the hadron momentum $\bf p$  and $\beta_i$ are the componets along  $\bf p,$  that is,  ${\bf k}_i {\bf p} = \beta_i {\bf p}^2.$  The hadron wave functions satisfy  the normalization condition  
\begin{eqnarray}
\sum_n\int\prod_{i=1}^n \frac{d\beta_id^2k_{\perp i}}{(2\pi )^2}\delta\left(1-\sum_{1=1}^n\beta_i\right)\delta^2\left(\sum_{1=1 }^nk_{\perp i}\right) \times \nonumber \\ \times
|\Psi\left( \beta_1, k_{\perp 1}; \beta_2, k_{\perp 2};\dots;\beta_n, k_{\perp n}\right)|^2 = 1.  \label{norm}
\end{eqnarray} 
The hadron wave function $\Psi\left( \beta_1, k_{\perp 1}; \beta_2, k_{\perp 2};\dots;\beta_n, k_{\perp n}\right)$  
satisfies  Schr\"odinger  equation  \cite{Lipatov:1974qm},
\begin{equation}
H\Psi = E(p)\Psi.  \label{Schro}
\end{equation}  
The quantity $E(p)$ appears in the energy propagators in the perturbation theory in the infinite hadron  momentum frame, in which  the energy propagators are 
\begin{eqnarray} 
 \left( E(p) - \sum_{1=1}^n E(k_i)\right)^{-1}, 
\end{eqnarray} 
here $E(k_i)$ is the energy of the parton.  Lipatov demonstrated  in \cite{Lipatov:1974qm} that this     Schr\"odinger  equation is equivalent to the DGLAP equation. From the solution to the DGLAP equation  we can obtain the solution for the    Schr\"odinger  equation for $n$ partons. 
In the next Sections we show an efficient method to solve DGLAP IDE by using Jacobians of the complex maps in the plane of the Mellin variable  \cite{Alvarez:2019eaa}. Numerically the Schr\"odinger equations may be solved  by the method of finite elements \cite{Kondrashuk:2015lix}.

\section{ Jacobians of complex maps}

In case when the coupling does not depend on the scale $u$ in Eq. (\ref{rge}), the solution of such an equation is just a power-like dependence on the variable $u.$  This happens in ${\cal N} =4$ supersymmetric Yang-Mills theory \cite{Alvarez:2016juq} and in Chern-Simons topological field theory \cite{Avdeev:1992jt}.  In such a  field theory 
an analog of the structure functions may be constructed, even if the asymptotic freedom is absent and the partons in this model would not have any physical sense \cite{Bianchi:2013sta}.  OPE in this model is valid for any momentum transfer $Q^2,$ that simplifies the calculations in the model, all the terminology used for "physical  quantities"  in  ${\cal N} =4$ supersymmetric Yang-Mills theory is based on  comparison with the case of QCD \cite{Alvarez:2016juq}.   Analogs of structure functions may be represented in the factorized form (\ref{domi}).
 
Anomalous dimensions are higly difficult to calculate them analytically in all the orders of the perturbation theory, however we  may always  consider the simplified case of the dominant "gluonic" contribution \cite{Alvarez:2016juq}.

To find the solution to the DGLAP equation we need to solve the differential equation  (\ref{rge})   for the Mellin moments and then to take the inverse Mellin transformation \cite{Alvarez:2016juq} .  The inverse Mellin transformation may be done by calculating the residues according to Cauchy integral formula, however this approach requires to use powerful software in the realistic case of higher orders in QCD which are necessary to compare with the experimental data \cite{Alvarez:2016juq}.  We propose to solve this task  in terms of Jacobians of complex maps in the complex plane of the Mellin moment variable $N$
 \cite{Alvarez:2019eaa}.
We take a toy model with the anomalous dimesion $\gamma(N) = 1/(N+1)$ and make these integral transformations  
\begin{eqnarray*} 
\phi(x,u) =  \int_{-1+\delta-i\infty}^{-1+\delta+i\infty}~dN \frac{x^{-N}}{N+1} u^{ \displaystyle{1/(N+1)}}   \\ 
= \int_{-1+\delta-i\infty}^{-1+\delta+i\infty}~dN \frac{1}{N+1}e^{-N\ln{x} + \ln{u}/(N+1)} \\
 = \oint_C dM  \frac{e^{M\sqrt{ \ln{u}\ln{\frac{1}{x}}}}  }{\sqrt{\le M+{1}/{w}\ri^2 - 4}}   \\ 
=  e^{-\sqrt{\ln{u}\ln{\frac{1}{x}}}/w} \oint_{C'} dM  \frac{e^{M\sqrt{ \ln{u}\ln{\frac{1}{x}}}}  }{\sqrt{ M^2 - 4}}  \\ = 
e^{-\ln{\frac{1}{x}}} \oint_{C''} dM  \frac{e^{2M\sqrt{ \ln{u}\ln{\frac{1}{x}}}}  }{\sqrt{ M^2 - 1}} = xI_0\le 2\sqrt{\ln{u}\ln{\frac{1}{x}}} \ri, 
\end{eqnarray*}
where $w \dis{\equiv \sqrt{ \frac{\ln{u}}{\ln{\frac{1}{x}}}}}.$ The complex map  
\begin{eqnarray}\label{map} 
N(M) = \frac{ Mw - 1 + \sqrt{(Mw + 1)^2 - 4w^2}}{2}
\end{eqnarray} 
to a new complex variable $M$ has been performed in the complex plane of the variable $N$ in order to represent the integral as a  Laplace transformation of the Jacobian of the corresponding complex map.   The integral contour $C''$ is an image of the straight line parallel to the imaginary axis after making several complex maps in the complex plane of the moment variable $N.$ It may be rectified to the 
straight vertical line again to apply a formula from the standard table of integrals or  can be curved to Hankel contour after destroying the factor 
$\sqrt{ M^2 - 1}$ of the Jacobian  by applying additional Mellin-Barnes transformation    \cite{Alvarez:2019eaa}.

Thus we have solution to the DGLAP equation (\ref{split}) in the form of the Bessel function  $xI_0\le 2\sqrt{\ln{u}\ln{\frac{1}{x}}}\ri.$  According to the Lipatov results   \cite{Lipatov:1974qm}  this DGLAP equation  (\ref{split}) is equivalent to  the Schr\"odinger equation (\ref{Schro}) for wave function of $n$ partons. This means that we may apply the method of complex maps to solve this equation of quantum mechanics.

\section{Equation dual to DGLAP}

The solution to the DGLAP equation (\ref{split}) represented  in the previous Section  in the form of the contour integral 
\begin{eqnarray} \label{orig}
 \phi(x,u) =  \int_{-1+\delta-i\infty}^{-1+\delta+i\infty}~dN \frac{x^{-N}}{N+1} u^{ \dis{ \frac{1}{N+1}}} 
\end{eqnarray}
may be re-written in the dual form by doing an appropriate complex map in the complex plane of the Mellin variable $N.$  The dual contour integral would bring the same 
Bessel function  $xI_0\le 2\sqrt{\ln{u}\ln{\frac{1}{x}}}\ri.$  

The new variable $M$  is related to the original variable $N$ by the duality relation $\gamma(N) = M. $ 
If we define the inverse function $\chi(M) \equiv N,$ then we model the duality condition  $\chi(\gamma(N)) =  N$ and $\gamma(\chi(M)) =  M.$  
The duality condition was used in the paper of Catani and Hautman in 1994 in \cite{Catani:1994sq}. The complex map based on this conditiion was performed in \cite{Kondrashuk:2019cwi}.

In this example (\ref{orig}) we may write explicitly (see \cite{Kondrashuk:2019cwi})
\begin{eqnarray*} 
\label{complex diffeo} \gamma(N) =  \frac{1}{N+1} = M \Rightarrow  N = \frac{1}{M} -1 \equiv \chi(M), \\  
\label{inverse} \chi(\gamma(N)) = \frac{1}{\gamma(N)} -1 = (N+1) -1 = N, \\
 \gamma(\chi(M)) = \frac{1}{\chi(M) +1} =  \frac{1}{\left(\displaystyle{\frac{1}{M}-1}\right)  +1} = M, 
\end{eqnarray*}
and make the corresponding diffeomorphism in the complex plane of the variable $N$
\begin{eqnarray*} 
\phi(x,u) =  \int_{-1+\delta-i\infty}^{-1+\delta+i\infty}~dN \frac{x^{-N}}{N+1} u^{ \displaystyle{ \frac{1}{N+1} }} \\
=  \oint_{C}~dM N'(M) \frac{x^{-\chi(M)}} {N+1} u^M = - \oint_{C}~dM \frac{x^{-\chi(M)} }{M} u^M 
\end{eqnarray*}
\begin{eqnarray*} 
= -\oint_{C}~dM \frac{x^{{1-1/M}}}{M} u^M =  - x\oint_{C}~dM \frac{x^{-1/M}}{M} u^M \\
 = - x\sum_{k=0}^{\infty}\frac{1}{k!}\le -\ln{x}\ri^k \oint_C~dM \frac{u^M}{M^{k+1}}  \\
= x\sum_{k=0}^{\infty}\frac{1}{k!}\le-\ln{x}\ri^k \frac{(\ln{u})^{k}}{k!} \\
= x\sum_{k=0}^{\infty}\frac{(-1)^{k}}{(k!)^2} \le\ln{u}\ln{x}\ri^k = xI_0\le 2\sqrt{\ln{u}\ln{\frac{1}{x}}} \ri.
\end{eqnarray*}
We have reproduced the same Bessel function, as it should be because any integral does not depend on any change of the variable. We may write the gluon unintegrated distribution function in a dual form 
\begin{eqnarray*} 
\phi(x,u) =  \oint_{C_1}~dN \phi(N,u) x^{-N} =    \oint_{C_2}~dM \phi(x,M) u^M \\
= xI_0\le 2\sqrt{\ln{u}\ln{\frac{1}{x}}} \ri.
\end{eqnarray*}
The contour $C_2$ is an image of the contour $C_1$ with respect to the mapping in the complex plane of the Mellin moment.  The functions $\phi(N,u)$ and $\phi(x,M)$ satisfy to the different differential equations, 
\begin{eqnarray*} 
u \frac{d}{d u} \phi \le N, u\ri   = \gamma(N,\alpha) \phi\le N, u\ri, \\
 x \frac{d}{d x} \phi \le x, M\ri   =  -\chi(M,\alpha) \phi\le x, M\ri. 
\end{eqnarray*}
The solutions to these equations are power-like functions, $\phi \le N, u \ri = \phi_1(N) u^{\gamma(N)}, ~~ \phi \le x, M \ri = \phi_2(M) x^{-\chi(M)},$ in which the functions  $\phi_1(N)$ and   $\phi_2(M)$ 
are related by the complex diffeomorphism (\ref{complex diffeo}) which is governed by the duality condition $\chi(\gamma(N)) =  N.$  After making the inverse Mellin transformation in the $N$ and $M$ planes, 
this couple of differential equations becomes a couple of dual IDEs, which are the DGLAP and the dual DGLAP  equations. This couple of IDEs is called dual because they contain the  same information.

\section{Regge limit for  dual DGLAP}

The dual DGLAP equation taken in the Regge limit under the duality  $\chi(\gamma(N)) =  N$ and $\gamma(\chi(M)) =  M$
may be treated as BFKL equation.  The BFKL equation is just a Regge limit of the optic theorem re-written in terms of the Bethe-Salpete equation \cite{Lipatov:1976zz,Fadin:1975cb,Kuraev:1976ge,Kuraev:1977fs,Balitsky:1978ic}.  The imaginary part of the forward scattering  amplitude 
(that is, total cross section of the deep inelastic scattering process)  is related to the correlators of two currents (\ref{W}),  too.  It is expectable that these two IDE (DGLAP and BFKL)  are related. The duality between  DGLAP equation and BFKL equation in QCD  was studied in  \cite{Altarelli:2000dw,Ball:2005mj}.  

In ${\cal N} =4$  supersymmetric Yang-Mills theory the structure of amplitudes is quite simple  and the Green functions are related to amplitudes in quite transparent way \cite{Kondrashuk:2004pu,Diaz:2024tdp}. In the real case of QCD the duality is investigated in the region of small $x$ \cite{Altarelli:2000dw,Ball:2005mj},  that is nearby only one point of the Riemann surface generated by the duality condition  $\gamma(N) = M. $    If we need to study dual DGLAP equation for more complicate case in comparison with considered here (\ref{orig}), for example in  ${\cal N} =4$  supersymmetric 
Yang-Mills theory,  without taking the Regge limit in all the dual DGLAP region, we need to take into account all the Riemann surface correspoding to the duality condition  $\gamma(N) = M$ \cite{Kondrashuk:2019cwi}.  In case of QCD the structure of the Riemann surface is even more sophisticated  due to the running of the coupling  \cite{Kondrashuk:2019cwi}.

\section{Dual DGLAP and unitarity}

It is not clear if the dual DGLAP equation, that is the IDE constructed from DGLAP equation via complex maps in the complex plane of the Mellin variable,  coincides with the optic theorem  completely but definitely the Regge limit of the dual DGLAP equation coincides with the Regge limit of the optic theorem  \cite{Kondrashuk:2019cwi}. 
This Regge limit of the optic theorem is called the BFKL equation.  This arises again very old question in quantum field  theory about relation 
between the unitarity (optic theorem in this case) and renormalization (DGLAP in this case). Apparently, the coincidence of the Regge limit of the dual DGLAP equation and 
Regge limit of the optic theorem suggests that these two in principle very different concepts of renormalization and of unitarity can be related 
via complex maps in  the plane of the Mellin moments, In turn, this means that the solution to the BFKL equation also may be found via Jacobians of the complex maps because we have shown  in   \cite{Alvarez:2019eaa} this method works well to solve the DGLAP equation.  

On the practical side, optic theorem may be applied to analyse tasks of quantum communication processes which may be solved by the methods 
of Jacobians of the complex maps considered here for the DGLAP and Schr\"odinger equations.

\section{Acknowledgments}
This work was supported by Fondecyt (Chile) grant 1050512 and by DIUBB (Chile) Grant Nos. 102609 and GI 153209/C.  I am very grateful to professor Vaclav Skala for  invitation to participate in this wondeful scientific event   "Quantum Informatics, Computing \& Technology 2025".

%-------------------------------------------------------------------------
% example of algorithm typesetting
% to allow this, uncomment line 
% \RequirePackage[noend]{myalgorithm}
% in the wscg.sty file
% and download that package from Gabriel Zachmann's page http://zach.in.tu-clausthal.de/latex/
%
%
%\begin{algorithm}
%\hrule
%  \centering
%\begin{algorithmic}
%    \STMT $d_{l,r} = f_B(P_1), f_B(P_n)$
%    \WHILE{ $|d_l| > \epsilon $ and $|d_r| > \epsilon $ and $l<r$}
%        \STMT $d_x = f_B(P_x)$
%        \IF{ $d_x < 0$ }
%            \STMT $l, r = x, r$
%        \ELSE
%            \STMT $l, r = l, x$
%        \ENDIF
%    \ENDWHILE
%\end{algorithmic}
%\hrule
%\caption{Example of some pseudo-code}
%\label{fg:code}
%\end{algorithm}

%-------------------------------------------------------------------------

\end{document}